\documentclass[10pt,letterpaper]{article}

\usepackage[margin=1in]{geometry}
\usepackage[T1]{fontenc}
\usepackage{lmodern}
\usepackage{microtype}
\usepackage{amsmath,amssymb,amsfonts}
\usepackage{graphicx}
\usepackage{textcomp}
\usepackage{tikz}
\usepackage{multirow}
\usetikzlibrary{positioning}
\usepackage{booktabs}
\usepackage{orcidlink}
\usepackage{cite}
\hypersetup{
  hidelinks,
  pdftitle={Reevaluating Bluetooth Low Energy for Ingestible Electronics},
  pdfauthor={Ziyao Zhou et al.},
  pdfsubject={Wireless communication for ingestible electronics}
}

\title{Reevaluating Bluetooth Low Energy for Ingestible Electronics}

\author{
\begin{tabular}{c}
Ziyao Zhou\orcidlink{0009-0007-0671-9947}, Zhuoran
Sun\orcidlink{0009-0007-2851-8599}, Xinyi
Shen\orcidlink{0009-0001-5608-6232}\\
Tianming
Zhou\orcidlink{0009-0009-1927-2691}, Yang
Li\orcidlink{0009-0005-8599-8248}, Rong
Tan\orcidlink{0000-0002-3207-7817}\\
Chen
Shen\orcidlink{0009-0008-4221-2534}, Tiancheng Cao\,\orcidlink{0000-0002-7259-5192}, Hen-Wei
Huang\orcidlink{0000-0003-1921-8897}
\end{tabular}
\thanks{This work was supported by the Nanyang Assistant Professorship, the MOE Tier 1 grants RG71/24 and RG26/25. \textit{(Corresponding author: Hen-Wei Huang.)}}
\thanks{Ziyao Zhou, Zhuoran Sun, Xinyi Shen, Tianming Zhou, Yang Li, Chen Shen, Tiancheng Cao, and Hen-Wei Huang are with the School of Electrical and Electronic Engineering, Nanyang Technological University, Singapore (e-mail: ZHOU0557, ZHUORAN004, SHEN0255, TIANMING003, YANG074@e.ntu.edu.sg; shenchen, tiancheng.cao, henwei.huang@ntu.edu.sg). Tiancheng Cao is also with Harvard Medical
School, Boston, MA, USA. Hen-Wei Huang is also with the Lee Kong Chian School of Medicine, Nanyang Technological University, Singapore.
}
\thanks{Rong Tan is with the Shenzhen Institutes of Advanced Technology, Chinese Academy of Sciences, Shenzhen, China (e-mail: rong.tan@siat.ac.cn).}
}
\date{}

\begin{document}
\maketitle

\begin{abstract}
Bluetooth Low Energy (BLE) has been widely adopted
in wearable devices; however, it has rarely been considered for ingestible electronics, primarily
due to concerns regarding severe tissue
attenuation at the 2.4 GHz band. In this work, we
systematically reevaluate the feasibility of BLE
for ingestible applications by benchmarking its
performance against representative Sub-GHz
communication schemes across tissue-induced
attenuation, power consumption, throughput,
communication reliability, bidirectional
communication capability, latency, and
system-level integration constraints.
Water-phantom experiments and in vivo measurements
consistently show stronger attenuation at 2.4 GHz
than at Sub-GHz bands, with attenuation of
approximately 40 dB; however, front-end-modules (FEM)-assisted BLE can
compensate for this additional loss and maintain a
practical received signal level while preserving energy efficiency. We quantify the
relationship between throughput and energy
consumption over a wide operating range and
demonstrate that, for the majority of ingestible
sensing applications with throughput requirements
below 100 kbps, FEM-assisted BLE with transmission power (TXP) of 20 dBm achieves substantially lower
power consumption than Sub-GHz alternatives. We
further show that BLE provides protocol-level
reliability through retransmission and in-order
packet delivery, while Sub-GHz links can
experience packet loss under high attenuation
conditions. In addition, BLE supports flexible
bidirectional communication with limited impact on
throughput, whereas the evaluated Sub-GHz
implementation exhibits a substantial throughput
reduction when bidirectional transmission is
enabled. End-to-end latency measurements show that
BLE offers lower latency than
Sub-GHz solutions due to its native compatibility
with modern computing infrastructure. Finally, we
analyze antenna form factor and ecosystem, highlighting the mechanical and
translational advantages of BLE in ingestible
system design. Collectively, these results
demonstrate that BLE, when properly
configured, provides a practical
communication solution for next-generation
ingestible electronics.

\end{abstract}

\noindent\textbf{Keywords:} BLE, Sub-GHz, wireless communication, ingestible electronics

\section{Introduction}


Wearable and medical Internet-of-Things (IoT)
technologies have undergone rapid translation into
clinical practice over the past decade, driven
largely by advances in low-power wireless
communication \cite{Habibzadeh2020A,
Huang2023Internet, Verma2022Internet}. Among
available short-range wireless standards,
Bluetooth Low Energy (BLE) has emerged as the
communication backbone for medical IoT, enabling
seamless connectivity between physiological
sensors, smartphones, bedside monitors, and cloud
infrastructure \cite{Rattal2025AIDriven,
Fama2022An}.

BLE is now widely adopted in wearable devices for
cardiac monitoring, activity tracking, and
home-based patient management, owing to its
ultra-low power operation, native smartphone
compatibility, mature security stack, and
standardized communication
profiles\cite{Tosi2017Performance}. However,
ingestible electronic devices designed to operate
within the gastrointestinal tract have
historically favored Sub-GHz radio frequency (RF)
communication, particularly in the 400–915 MHz
bands \cite{Abdigazy2024End-to-end}. This
preference is primarily motivated by the lower
electromagnetic attenuation of biological tissues
at Sub-GHz frequencies, which provides a superior
link margin for in-body communication. As a
result, BLE has often been dismissed as unsuitable
for ingestible systems due to its operation in the
2.4 GHz band, where tissue absorption and
multipath loss are more pronounced
\cite{Christoe2021Bluetooth}.

\begin{figure}[htbp]
    \centering
    \includegraphics[width=\columnwidth]{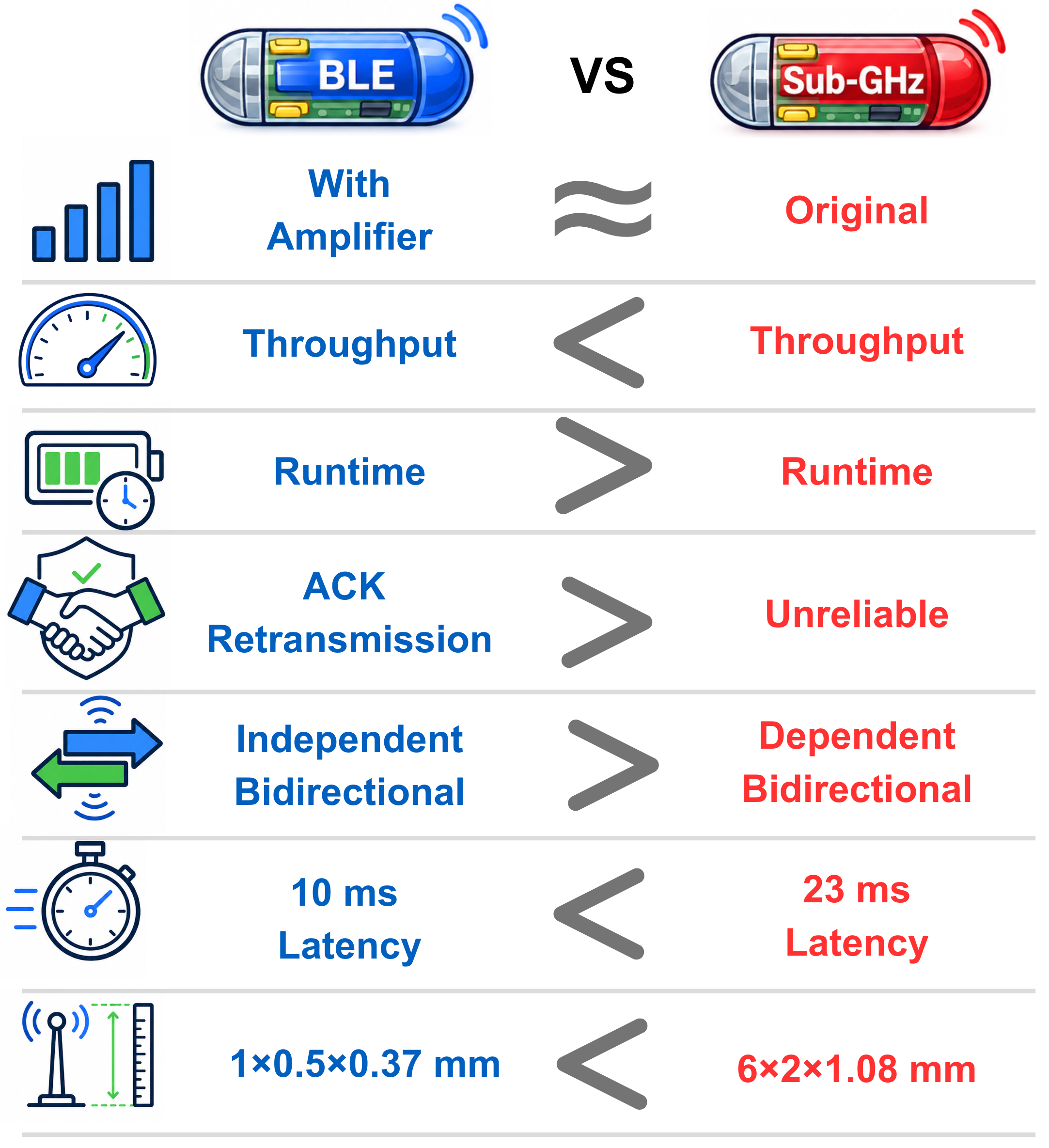}
    \caption{Summary comparison between BLE and Sub-GHz communication for ingestible electronics in terms of signal strength, throughput, power consumption, reliability, bidirectional transmission, end-to-end latency, and antenna size.}
    \label{fig:overview}
\end{figure}

This frequency-centric view overlooks several
critical considerations for ingestible
electronics, as illustrated in Fig.~\ref{fig:overview}.
First, BLE is explicitly designed for
energy-constrained applications and typically
consumes less power than Sub-GHz
alternatives\cite{Morin2017Comparison}. Second,
attenuation does not necessarily indicate
insufficient throughput, and throughput
requirements vary by application, which must be
assessed in context. Third, even in scenarios
where BLE experiences significant attenuation, its
performance can be compensated by incorporating
front-end modules (FEM) to boost transmission
power (TXP)\cite{Zhou2025Closed-Loop}.
In such cases where throughput requirements is not high, BLE with FEM may still offer lower
power consumption compared to Sub-GHz systems.

Communication reliability and bidirectional
capability are also critical for ingestible
electronics. In vivo sensing data must be
delivered with minimal data loss, especially when
the transmitted information is used for
physiological monitoring, closed-loop control, or
clinical decision-making \cite{Kim2023Advances, Ghanim2023Communication}. Meanwhile, many
ingestible systems require downlink commands from
the external receiver to the in-body device for
device configuration, adaptive sensing, actuation,
or controlled drug release \cite{Zhang2025Ingestible}. BLE inherently
supports reliable packet delivery and
bidirectional data exchange at the protocol level,
whereas many Sub-GHz modules do not provide
comparable protocol-level support. In addition,
modules operating at different frequencies often
require distinct peripheral components, which can
influence the mechanical design and form factor of
ingestible devices \cite{Abdigazy2024End-to-end}.

In addition, BLE offers significant system-level
advantages that are highly relevant to ingestible
electronics. It supports native interoperability
with consumer smartphones and hospital IT
infrastructure, eliminating the need for custom
receivers or proprietary gateways
\cite{Swaroop2019A}. This reduces deployment complexity, improves patient compliance, and
accelerates regulatory and clinical translation.
Moreover, BLE integrates mature features such as
encryption, device authentication, and
over-the-air firmware updates—capabilities that
often require custom implementation or are less
standardized in Sub-GHz platforms, yet are
critical for cybersecurity and medical device
lifecycle management \cite{Barua2022Security,
Zheng2025Research}.

Taken together, these observations motivate a
re-examination of BLE as a viable wireless
communication strategy for ingestible electronics.
We first characterize tissue-induced attenuation
at 433 MHz, 915 MHz, and 2.4 GHz using
water-phantom experiments and in vivo gastric
measurements. Next, we evaluate the relationship
between throughput and power consumption across
configurations under representative throughput
demands. We further assess communication
reliability by measuring packet loss under
different Received Signal Strength Indicator
(RSSI) conditions and evaluate bidirectional
communication capability by quantifying the
trade-off between forward and reverse throughput.
We then quantify the latency of BLE and the
Sub-GHz configurations. Finally, we highlight the
advantages of BLE over Sub-GHz technologies in
terms of mechanical integration and ecosystem
compatibility. Overall, our findings show that
properly configured BLE can provide a robust and
scalable wireless platform for next-generation
ingestible electronics.


\section{Methodology}


\begin{table}[htbp]
\centering
\caption{Sub-GHz communication module comparison.}
\label{tab:subghz_comparison}
\footnotesize
\setlength{\tabcolsep}{2pt}
\renewcommand{\arraystretch}{1.08}
\resizebox{\textwidth}{!}{%
\begin{tabular}{llcccccc}
\hline
\textbf{Module} &
\textbf{Vendor} &
\textbf{Freq. Range (MHz)} &
\textbf{Max TXP (dBm)} &
\textbf{Max Data Rate} &
\textbf{TX Current (mA)} &
\textbf{RX Current (mA)} &
\textbf{Estimated TX Energy (nJ/bit)} \\
\hline
MM8108 &
Morse Micro &
850--950 &
+25.5 &
43.3 Mbps 802.11ah &
107 @ +16 dBm\textsuperscript{\dag} &
34\textsuperscript{\dag} &
8.2 \\

ZL70323 &
Microchip &
402--405 (MICS) &
$-$3.5 &
200 kbps FSK &
5.6 @-3.5 dBm &
4.3 &
84.0 \\

SX1261 &
Semtech &
150--960 &
+15 &
500 kbps GFSK / 62.5 kbps LoRa &
32.5 @ +15 dBm &
4.8 (GFSK) / 5.3 (LoRa) &
214.5 (GFSK) / 1716 (LoRa) \\

Si4461 &
Silicon Labs &
142--1050 &
+16 &
1 Mbps GFSK &
43 @ +16 dBm &
10.9--13.7 &
141.9 \\ 

CC1310 &
Texas Instruments &
287--1054 &
+15 &
4 Mbps GFSK &
23.5 @ +14 dBm &
5.5 &
21.2 \\

\hline
\multicolumn{8}{l}{\textsuperscript{\dag} At 8 MHz channel bandwidth, MCS9.} \\
\end{tabular}%
}
\end{table}

To select a representative Sub-GHz module for
subsequent comparison with BLE, we first screened
several commercially available Sub-GHz
communication modules with different frequency
ranges, modulation schemes, and design targets.
Table~\ref{tab:subghz_comparison} summarizes their
operating frequency, maximum TXP, data rate,
current consumption, and estimated TX energy per
bit. The comparison shows that these modules are
optimized for different communication scenarios,
leading to clear trade-offs among throughput, peak
current, energy efficiency, and application
constraints.

The MM8108 achieves the highest data rate among
the evaluated modules, reaching 43.3 Mbps under
802.11ah Wi-Fi HaLow
\cite{MM8108_datasheet}. Its estimated TX energy per bit is the
lowest, at 8.2 nJ/bit, primarily due to its high
throughput. However, this advantage is accompanied
by a high TX current of 107 mA, which may impose a
substantial instantaneous power burden on
power-constrained ingestible systems. Therefore,
although MM8108 is attractive for burst-mode
high-throughput transmission, it is less suitable for general low-power ingestible applications

The ZL70323 represents a dedicated implantable
communication solution operating in the MICS band.
It has a low TX current of 5.6 mA, but its maximum
TXP is restricted to -3.5 dBm, and its data rate
is limited to 200 kbps \cite{ZL70323_datasheet}.
These features make it suitable for low-data-rate
implantable telemetry, but its limited link budget
and throughput restrict its applicability to
broader ingestible applications that require
higher data transmission capability.

The SX1261 and Si4461 provide wide frequency
coverage and comparable maximum TXP levels, but
their relatively lower data rates result in higher
estimated TX energy per bit. This limitation is
particularly evident for the SX1261 in LoRa mode,
where the energy cost increases to 1716 nJ/bit
because of the reduced data rate. Although these
modules can support robust or long-range Sub-GHz
links, they are not the most efficient options
when moderate-to-high energy efficiency is
required.

In contrast, the CC1310 offers the most balanced
performance among the evaluated Sub-GHz modules.
It supports a relatively high data rate of 4 Mbps
while maintaining a moderate TX current of 23.5 mA
at +14 dBm \cite{CC1310_datasheet}, yielding an
estimated TX energy of 21.2 nJ/bit. Compared with
the SX1261, Si4461, and ZL70323, the CC1310
provides higher throughput and lower energy cost
per bit. Compared with the MM8108, it requires
substantially lower TX current while still
providing sufficient throughput for most
sensing-oriented ingestible applications.
Therefore, the CC1310 was selected as the
representative Sub-GHz module in this study and
evaluated at two frequency bands, 433 MHz and 915
MHz, both with a fixed TXP of 8 dBm. For the 433
MHz configuration, the highest available 2-FSK
modulation setting was used. For the 915 MHz
configuration, the high-speed (4 Mbps) mode based
on 8-FSK modulation was adopted to achieve
higher-throughput Sub-GHz communication.

For the BLE platform, we selected the nRF54L15 to
provide a representative low-power 2.4 GHz
implementation. The nRF54L15 is Nordic
Semiconductor's latest-generation ultra-low-power
2.4 GHz radio SoC, supporting BLE physical layer (PHY) modes from
Coded PHY (S8) to 2 Mbps and a maximum on-chip TXP
of 8 dBm. In all BLE experiments, the connection
interval was fixed at 7.5 ms. Two experimental
configurations were evaluated. The first used the
standalone nRF54L15 operating at a fixed TXP of 8
dBm, whereas the second integrated an external
nRF21540 front-end module (FEM) with the nRF54L15,
extending the maximum effective TXP to 20 dBm.
During the experiment, BLE and the CC1310
Sub-GHz link were configured with their maximum
packet payload lengths, corresponding to 244
bytes for BLE and 254 bytes for the CC1310
Sub-GHz configuration.

Power consumption during both BLE and Sub-GHz
communication was measured using the Nordic Power
Profiler Kit II (PPK), configured as a 1.8 V power
supply and current-measurement instrument for the
transmitter. The PPK provides high-resolution
current profiling at a sampling rate of 100,000
samples per second
\cite{nordicsemi_ppk2_userguide_v1.0.1}.

\subsection{Tissue-Mimicking Penetration Evaluation}
\begin{figure}[htbp]
    \centering
    \includegraphics[width=\linewidth]{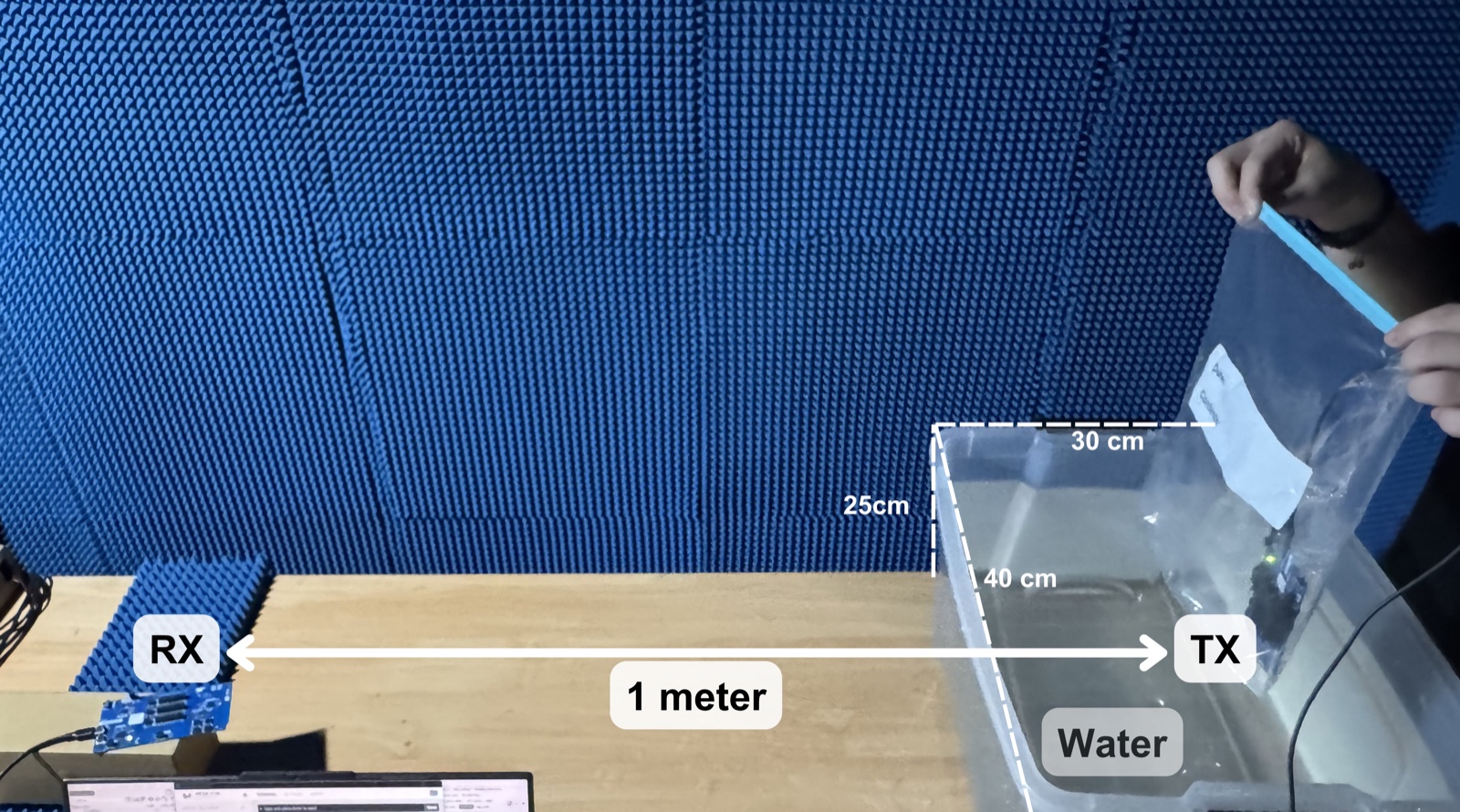}
    \caption{Experimental setup for evaluating communication performance through a water-based tissue phantom in a microwave anechoic chamber.}
    \label{fig:anechoic_chamber}
\end{figure}




To emulate human tissue attenuation, water was
used as the propagation phantom in a rectangular
water tank. This choice was motivated by the fact
that approximately 70\% of the human body consists
of water and that water exhibits dielectric
properties comparable to those of biological
tissues
\cite{Nascimento2020Comparative}. The experimental setup is shown in Fig.
\ref{fig:anechoic_chamber}.

Measurements were conducted at six immersion
depths, with the transmitter positioned at the
center of the water tank at depths of 0, 2, 4, 6,
8, and 10 cm. The transmitter was submerged in
water, while the receiver was positioned in the
air at a fixed distance of 1 m. First, a target
throughput of 50 kbps was selected, as it is
sufficient to meet the requirements of the vast
majority of practical ingestible sensing
applications. Subsequently, each communication
configuration was operated at its maximum
achievable throughput to evaluate attenuation
effects in high-throughput applications. For each
immersion depth and communication configuration,
the transmitter power consumption, receiver
throughput, and RSSI were recorded.
 
\subsection{In Vivo Signal Evaluation}

Existing studies have experimentally investigated
Sub-GHz signal propagation within the
gastrointestinal tract. For example, in
\cite{9363615}, the transmitting antenna was placed at the bottom
of the bovine rumen, and the attenuation
characteristics of Sub-GHz signals at 433 and 868
MHz were evaluated.
\cite{Say2026Bioresorbable} further reported the RSSI attenuation
of a 915 MHz RFID capsule in air, in an ex vivo
porcine stomach, and in an in vivo porcine
stomach. However, for 2.4 GHz communication,
specific experimental attenuation data within the
gut remain limited. To the best of our
knowledge, only measured path-loss data at 2.4 GHz
in the abdominal cavity, rather than inside the
gastrointestinal tract, have been reported \cite{9849637}.

\begin{figure}[htbp]
    \centering
    \includegraphics[width=\columnwidth]{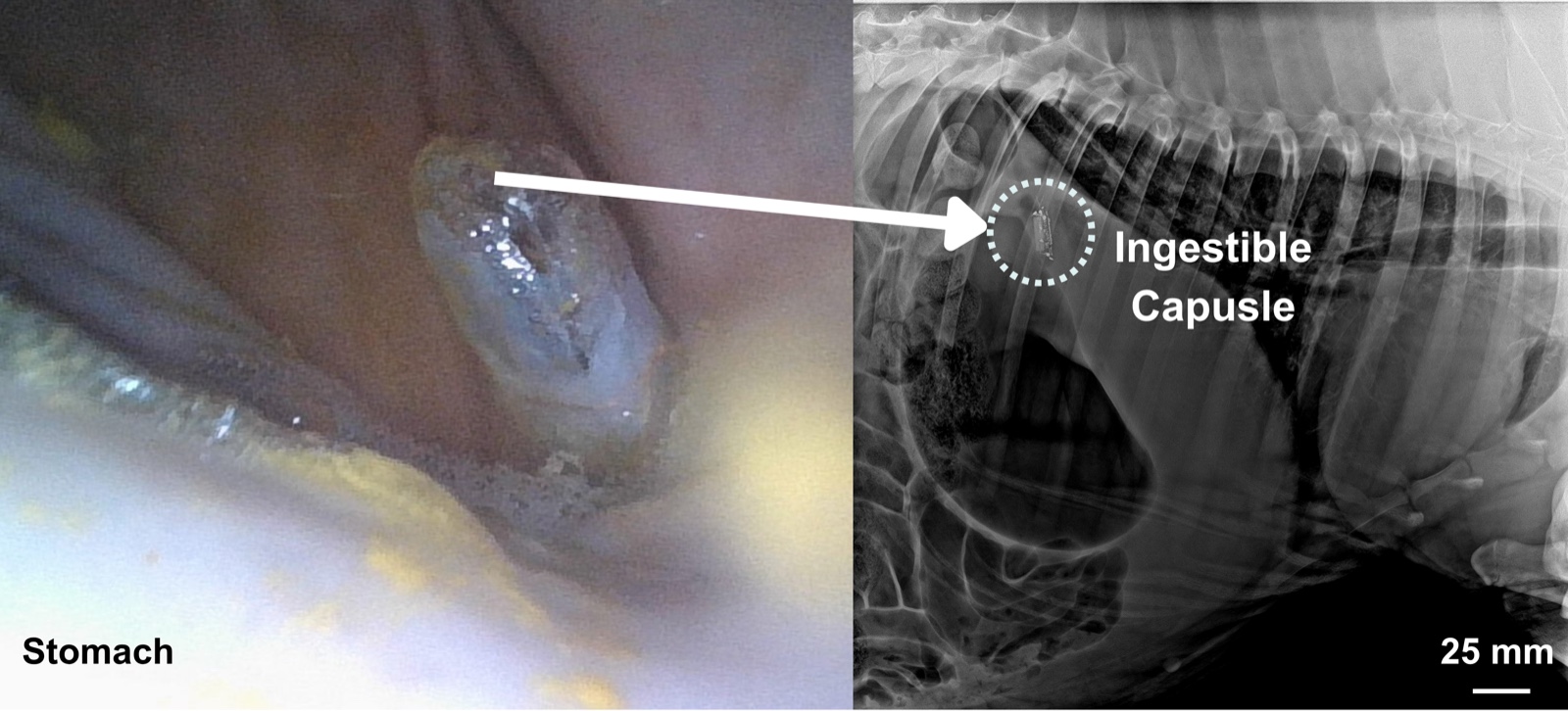}
    \caption{In vivo validation of the 2.4 GHz BLE ingestible capsule location in the stomach.}
    \label{fig:xray}
\end{figure}

To address this gap and quantify the in vivo
attenuation of 2.4 GHz BLE signals, we designed a
BLE capsule
\cite{Huang2026Kirigami}. Using a method similar to that shown in
Fig.
\ref{fig:anechoic_chamber}, we first measured the RSSI of the capsule in air. The capsule was then
delivered into the porcine stomach, as confirmed
by X-ray and endoscopic images shown in Fig. \ref{fig:xray}. The
difference in RSSI before and after ingestion was
used to quantify tissue-induced signal attenuation
at 2.4 GHz.

\subsection{Power--Throughput Evaluation}

The choice of wireless communication modules
should be guided by the throughput requirements of
the target application, ensuring reliable data
delivery and stable links. Ingestible electronics
span a wide range of data throughput requirements,
typically from a few bits per second (bps) to
hundreds of kilobits per second (kbps), depending
on the sensing modality. Slow-varying
physiological signals—including pH, temperature,
and pressure—generally require around 10–200 bps
\cite{Maqbool2009SmartPill}. Electrochemical and gas sensors also fall within this range
\cite{Kalantar-Zadeh2018A}. Monitoring of basic vital signs, including respiratory and heart rate,
requires moderate throughput around 2 kbps
\cite{Traverso2023First-in-human}. Electrogastrography (EGG) may
demand higher throughput due to the large data
volume across multiple ADC channels, often
reaching tens of kbps
\cite{You2024An}. Notably, except for capsule endoscopy and
heart-sound monitoring, which can require
substantially higher throughputs of up to several
Mbps
\cite{Ghanim2023Communication}, most ingestible applications have relatively low throughput
requirements, generally below 100 kbps. These
observations underscore the need to evaluate
wireless communication strategies for ingestible
devices on an application-specific basis.

To evaluate power consumption under different
throughput levels, both BLE and Sub-GHz systems
were configured by adjusting the data transmission
duty cycle to achieve the desired throughput,
ranging from 50 bps to 2300 kbps. Four
communication configurations were evaluated,
consistent with the previous experiment: nRF54L15
operating at 8 dBm, nRF54L15 operating at 20 dBm,
CC1310 operating at 433 MHz, and CC1310 operating
at 915 MHz. The transmitter was placed at the 0 cm
position in the tissue phantom, while the receiver
position remained unchanged from the previous
experiment.

\subsection{Communication Reliability Evaluation}

Communication reliability is critical for
ingestible electronics because the transmitted
payload often contains time-sensitive in vivo data
that are difficult to reacquire after the device
has been deployed. Packet loss, data corruption,
or out-of-order delivery may interrupt continuous
sensing records and reduce the reliability of
feedback control. Therefore, reliable packet
delivery is an important requirement in addition
to throughput and power consumption.

To compare the communication reliability of BLE
and Sub-GHz systems, each transmitter was
configured to send sequentially numbered data
packets, while the receiver decoded the packet
sequence numbers and counted the number of missing
packets. During the experiment, a tunable RF
attenuator was used to introduce controlled
attenuation levels between the transmitter and
receiver. The packet loss rate was then calculated
under each RSSI condition to characterize
communication reliability. For BLE, in addition to
the baseline 2 Mbps PHY, the Coded PHY (S8) was
also evaluated to verify its reliability-enhancing
capability under attenuated link conditions.

\subsection{Bidirectional Communication Evaluation}

In ingestible electronics, communication is not
limited to forward transmission from the in-body
device to the external receiver. Reverse
transmission from the external receiver to the
in-body device is also frequently required for
tasks such as device configuration, sensing-mode
adjustment, and adaptive data acquisition. BLE
natively supports bidirectional communication by
sharing the radio resource through half-duplex
time-division multiplexing. In this architecture,
either side of the BLE link can independently
initiate data transmission within the connection
events, enabling flexible two-way data exchange
between the ingestible device and the external
host
\cite{BluetoothCoreSpecV62}.

\begin{figure}[htbp]
\centering
\begin{tikzpicture}[
    >=stealth,
    font=\footnotesize,
    state/.style={
        rectangle, rounded corners=3pt,
        draw=black, semithick,
        minimum width=1.9cm, minimum height=0.58cm,
        align=center, inner sep=4pt
    },
    trans/.style={->, semithick},
    rf/.style={->, semithick, dashed}
]

\node[state, fill=gray!12] (I1) {Send DATA};
\node[state, fill=gray!12, right=1.3cm of I1] (I2) {Await ACK};

\node[state, fill=white, below=1.7cm of I1] (R1) {Await DATA};
\node[state, fill=white, right=1.3cm of R1]  (R2) {Send ACK};

\node[font=\scriptsize\bfseries, left=0.18cm of I1] {Initiator};
\node[font=\scriptsize\bfseries, left=0.18cm of R1] {Responder};

\draw[trans] (I1) -- node[above]{sent} (I2);
\draw[trans] (I2) to[bend left=38]
    node[below]{\makebox[0pt]{ACK / timeout}} (I1);

\draw[trans] (R1) -- node[above]{received} (R2);
\draw[trans] (R2) to[bend left=38] node[below]{sent} (R1);

\draw[rf] (I1.south) -- node[left]{DATA} (R1.north);
\draw[rf] (R2.north) -- node[right]{ACK}  (I2.south);

\end{tikzpicture}
\caption{State machine of the proposed real-time bidirectional Sub-GHz link.
  The initiator cycles between transmitting a DATA frame and awaiting
  the acknowledgment; the responder cycles between awaiting DATA and
  returning the ACK. Dashed arrows denote over-the-air transmissions.}
\label{fig:bidirectional_sm}
\end{figure}

Sub-GHz operation does not inherently define a
common link-layer acknowledgment or bidirectional
scheduling mechanism; instead, these functions
depend on the specific protocol stack and firmware
implementation. In this study, the CC1310 4M PHY provided
unidirectional packet transmission over a
proprietary PHY but did not include acknowledgment,
retransmission, or bidirectional scheduling.
Therefore, we extended the firmware with a custom
software-based stop-and-wait protocol, implemented
using two alternating states at each endpoint, as
illustrated in Fig.~\ref{fig:bidirectional_sm}.
The initiator transmitted one DATA frame, switched
to receive mode, and waited for a response within
a configurable timeout window. Upon successfully
decoding the DATA frame, the responder switched
from receive to transmit mode and returned a
response frame before resuming reception. This
response frame both acknowledged reception of the
DATA frame and carried a configurable reverse
payload. The reverse-payload length was varied to
quantify the trade-off between forward and reverse
throughput. Because each exchange required radio
turnaround and response-frame transmission, this
implementation incurred additional TX/RX switching
latency and protocol overhead.

The forward throughput of both the BLE and Sub-GHz
systems was varied. Under each forward-throughput
setting, the maximum achievable reverse throughput
was measured. Therefore, the resulting data
represent the boundary of bidirectional
communication capability for each wireless module.
\begin{figure}[htbp]
    \centering
    \includegraphics[width=\textwidth]{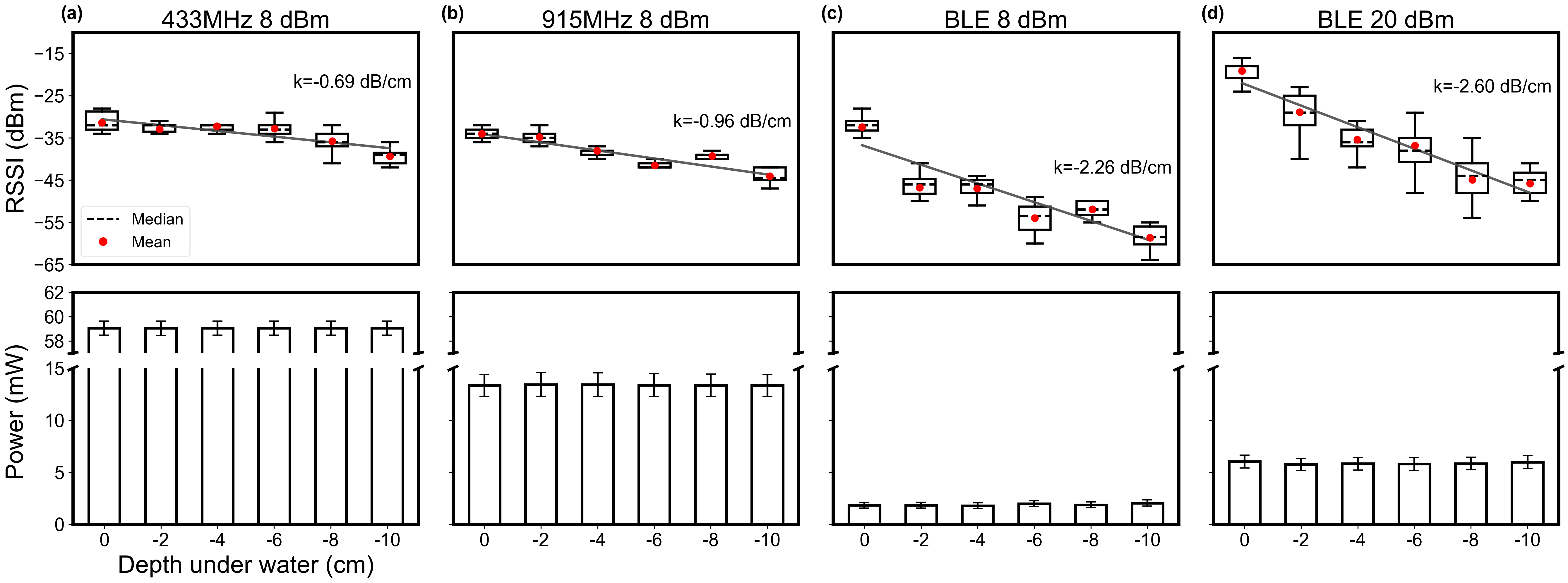}
    \caption{Comparison of Sub-GHz communication at (a) 433 MHz and (b) 915 MHz, as well as BLE with TXP of (c) 8 dBm and (d) 20 dBm, in terms of RSSI and power consumption at various water depths under a 50 kbps transmission condition.}
    \label{fig:Transmission characteristics}
\end{figure}

\begin{figure}[htbp]
    \centering
    \includegraphics[width=0.9\columnwidth]{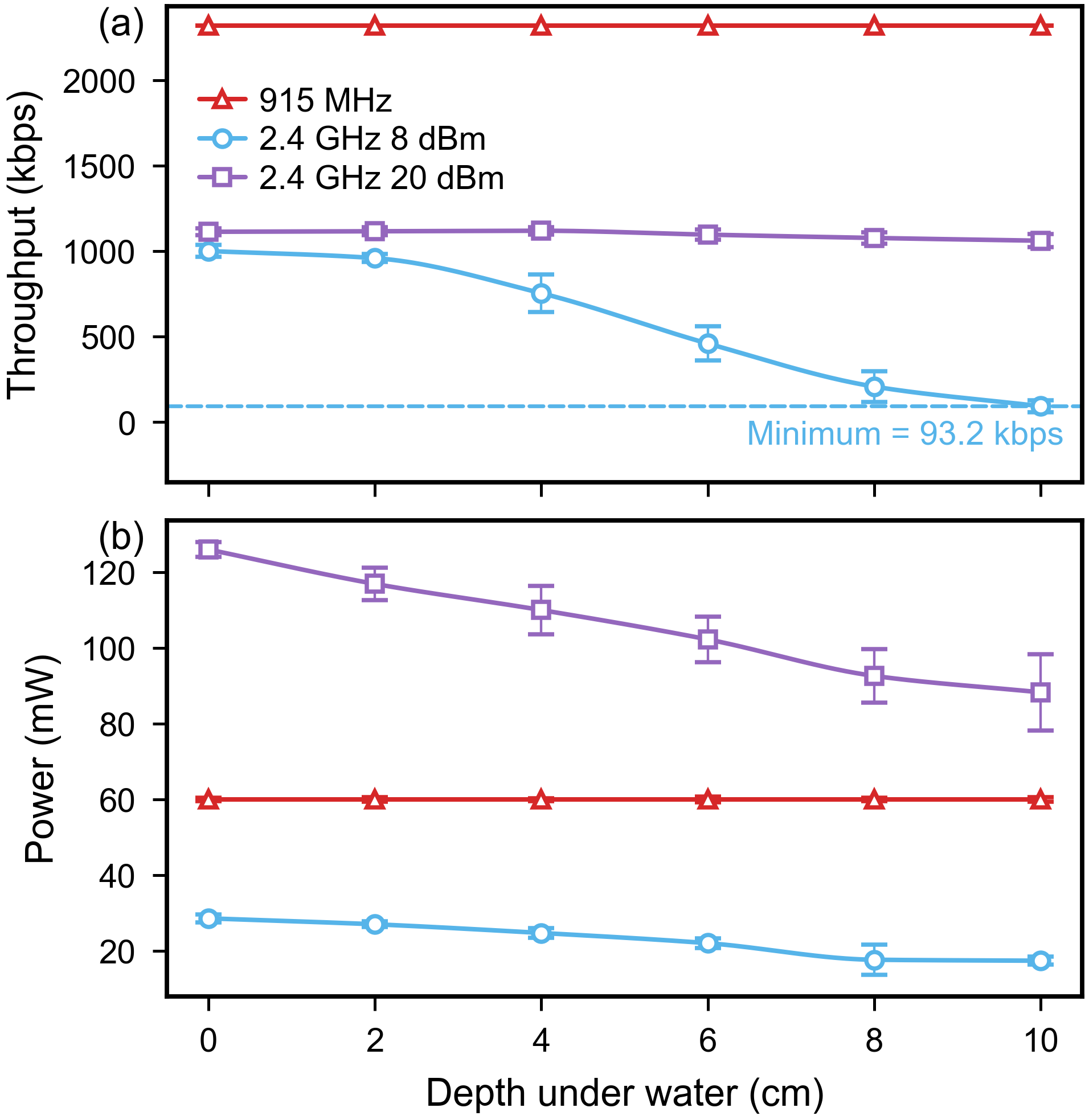}
    \caption{Maximum-throughput performance of BLE and Sub-GHz communication links under different water depths. (a) Throughput variation with water depth. (b) Power consumption variation with water depth.}
    \label{fig:water_attenuation}
\end{figure}

\subsection{End-to-End Latency Evaluation}
Medical devices are highly sensitive to
communication latency, and different applications
impose different requirements on low-latency
performance. In general, lower communication
latency enhances data timeliness and improves
clinical effectiveness. Life-critical
applications, such as surgery, require end-to-end
delays below 10 ms \cite{10388245}. As a result,
these systems are predominantly implemented using
wired connections to ensure deterministic and
ultra-low latency. In comparison, vital sign
monitoring—such as ECG, EEG, EMG, temperature,
blood pressure, and respiratory rate—can tolerate
latencies up to 250 ms\cite{10388245}. While
increased latency in these systems may degrade
clinical value or temporal resolution, it does not
typically pose an immediate risk to patient
safety.

BLE theoretically offers advantages in
communication latency, as most hospital IT
infrastructures are already equipped with network
interfaces that can directly communicate with BLE
devices. To evaluate the practical board-to-board
and end-to-end latency, the transmitter
periodically sent 244-byte packets. Board-to-board
latency quantifies the intrinsic BLE link delay,
whereas end-to-end latency includes additional
system-level overhead from data reception and
processing on the computing platform. Another
development board and an NVIDIA Jetson Nano served
as BLE receivers for board-to-board and end-to-end
measurements, respectively. For one-way latency
measurement, a GPIO pulse was generated on the
transmitter at the start of each transmission, and
another GPIO pulse was generated on the receiver
upon successful reception of the complete packet.
The time difference between the two pulses was
recorded using an oscilloscope (RIGOL MSO5074).
Two-way latency was measured by additionally
requiring the receiver to immediately echo the
received packet back to the transmitter, with a
second GPIO pulse generated on the transmitter
upon reception of the echoed data.

The same board-to-board and end-to-end latency
measurements were performed for Sub-GHz
communication.  In contrast to BLE, Sub-GHz
communication is expected to exhibit higher
end-to-end latency because general-purpose
computing devices cannot directly interface with
Sub-GHz radios and therefore require an external
USB-connected transceiver. For board-to-board
measurement, a CC1310 development board served as
the receiver, whereas for end-to-end measurement,
the NVIDIA Jetson Nano received data through the
USB-connected Sub-GHz transceiver. A GPIO pulse
was generated on the CC1310 transmitter upon
completion of each 244-byte frame transmission,
and a second pulse was generated on the receiver
once the frame had been fully received. One-way
latency was obtained from the time difference
between these two pulses. For two-way latency, the
receiver immediately relayed the received frame
back to the transmitter via the Sub-GHz link, and
an additional GPIO pulse on the transmitter marked
the receipt of the echoed data.

\section{Results and Discussion}


\subsection{Tissue-Mimicking Penetration Evaluation}

The transmission characteristics of the four
communication configurations at six underwater
depths are shown in Fig.
\ref{fig:Transmission characteristics}. Under the 50 kbps
transmission condition, all four configurations
are able to sustain stable communication across
all tested water depths, indicating that both BLE
and Sub-GHz links can satisfy the requirement of
low-throughput ingestible sensing applications.

The RSSI measurements show clear
frequency-dependent attenuation across different
water depths. For the 433 MHz Sub-GHz
configuration, the RSSI decreases only mildly
across different water depths, with an overall
reduction of approximately 10 dB. This
demonstrates the superior penetration capability
of lower-frequency signals. For the 915 MHz
configuration, the RSSI indicates stronger
attenuation than that at 433 MHz, with an overall
RSSI reduction of approximately 20 dB. For BLE
operating at 8 dBm, the overall RSSI is lower and
exhibits a steep attenuation. BLE operating at 20
dBm shows a similar attenuation slope, but with
the additional 12 dB of TXP from the external
amplifier; thus, its absolute RSSI remains higher,
resulting in an improved signal-to-noise ratio at
the receiver. Overall, the attenuation of BLE
signals is approximately 40 dB.

In terms of power consumption, all four
communication configurations exhibit nearly
constant power levels across different water
depths. For BLE, operation at 20 dBm results in
approximately three times the power consumption of
the 8 dBm configuration; however, the introduction
of the external amplifier significantly improves
channel quality. In comparison, the power
consumption of the 915 MHz Sub-GHz configuration
is approximately three times that of BLE operating
at 20 dBm, while the 433 MHz configuration
consumes nearly 10 times more power. Therefore,
for applications with low or moderate throughput
requirements, incorporating an amplifier into BLE
systems provides a practical means to ensure
reliable communication while achieving
substantially lower power consumption than Sub-GHz
alternatives.

However, in high-bandwidth applications such as
capsule endoscopy, BLE faces a trade-off between
attenuation robustness and power consumption. As
shown in Fig. \ref{fig:water_attenuation}, the
throughput of BLE without an FEM decreases with
increasing water depth and reaches a minimum of
93.2 kbps at the largest attenuation condition.
Although this value is insufficient for
high-throughput applications such as capsule
endoscopy, it remains above the 50 kbps
requirement of low-throughput ingestible sensing
applications. In contrast, FEM-assisted BLE
maintains a higher throughput of approximately
1100 kbps, but its power consumption remains
around 100 mW. The CC1310 operating at 915 MHz
consistently achieves approximately 2300 kbps
across all tested water depths, with no obvious
attenuation and a power consumption of
approximately 60 mW. The 433 MHz configuration is
not included in this comparison because its
maximum achievable throughput is limited to
approximately 50 kbps, far below the requirement
of high-throughput applications.

\subsection{In Vivo Signal Evaluation}

As summarized in Table
\ref{tab:in_vivo_attenuation}, the 433 MHz band
reported in vivo attenuation is 23.20 dB, which
differs from our water-phantom experiment, where
433 MHz exhibits the smallest attenuation of
approximately only 10 dB. This discrepancy may be
attributed to the use of an Electromagnetic Fields
(EMF) probe rather than a dedicated 433 MHz
antenna in \cite{9363615}, which could increase
the apparent attenuation compared with an
optimized antenna-based receiver.

For the 915 MHz band, the measured in vivo
attenuation is 16.85 dB. This result follows the
same trend observed in our water-phantom
experiment, where the overall RSSI reduction is
approximately 20 dB. Both evaluations consistently
indicate that 915 MHz provides a smaller link
margin than 433 MHz for in-body transmission.

2.4 GHz shows the highest attenuation among the
three bands, confirming that BLE experiences
stronger tissue-induced loss in the gastric
environment. The measured in vivo attenuation is
consistent with the water-phantom experiment,
which shows an overall RSSI reduction of
approximately 40 dB. This comparable attenuation
level supports the relevance of the water-phantom
results and confirms that FEM-assisted BLE can
effectively compensate for the higher in vivo
tissue-induced loss at 2.4 GHz.

\begin{table}[htbp]
  \centering
  \caption{Measured in vivo signal attenuation at different frequencies.}
  \label{tab:in_vivo_attenuation}
  \resizebox{\columnwidth}{!}{%
  \begin{tabular}{@{}cccc@{}}
    \toprule
    Frequency & Mean Attenuation (dB) & Standard Deviation (dB) & Ref. \\
    \midrule
    433 MHz & $23.20$ & $3.80$ & \cite{9363615} \\
    915 MHz & $16.85$ & $2.95$ & \cite{Say2026Bioresorbable} \\
    2.4 GHz & $42.85$ & $6.06$ & This work \\
    \bottomrule
  \end{tabular}%
  }
\end{table}

The use of high radio-frequency transmission power in ingestible electronics inevitably raises safety concerns because excessive RF exposure may cause localized energy absorption in surrounding tissues. Therefore, although the 20 dBm FEM-assisted BLE configuration compensates for the higher tissue attenuation at 2.4 GHz, its safety must be evaluated based on the specific absorption rate (SAR) under the intended operating conditions. IEEE C95.1-2019 specifies a 10 g averaged SAR limit of 2 W/kg for localized exposure \cite{9238523}. A previous study of a capsule-conformal antenna at 2.45 GHz reported a SAR-derived maximum antenna-port input power of approximately 19 dBm under continuous excitation \cite{9667256}. In the present system, however, the specified 20 dBm represents the chip/FEM output power rather than the power delivered to the antenna. Matching, interconnect, and packaging losses reduce the actual antenna-port input power. Moreover, the low-throughput sensing applications considered in this work use duty-cycled transmission, resulting in a time-averaged delivered power substantially below the peak FEM output. These factors suggest that the 20 dBm BLE configuration is unlikely to exceed the 10 g averaged SAR limit under the intended duty-cycled operating conditions. Nevertheless, device-specific SAR simulation or measurement incorporating the final antenna, packaging, and transmission duty cycle is required to verify compliance.

\subsection{Power--Throughput Evaluation}

\begin{figure}[htbp]
    \centering
    \includegraphics[width=0.95\textwidth]{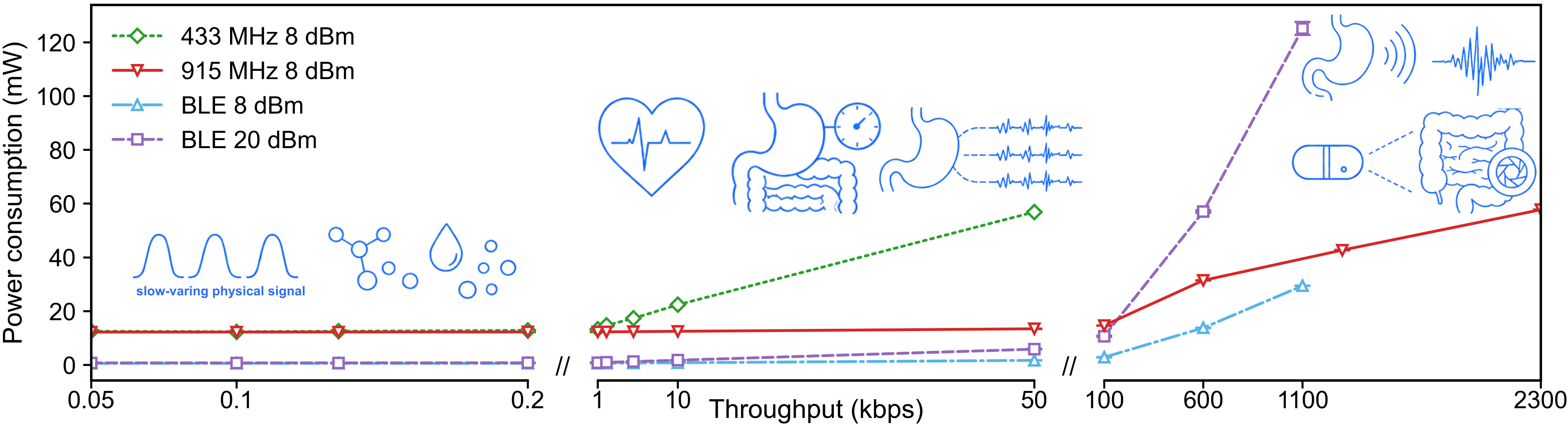}
    \caption{Comparison of power consumption versus throughput for BLE and Sub-GHz communication systems.}
    \label{fig:power_vs_throughput}
\end{figure}

The relationship between throughput and power
consumption under the four communication
configurations is shown in Fig.
\ref{fig:power_vs_throughput}. Comparing BLE operating at 8 dBm
and 20 dBm with the 433 MHz Sub-GHz configuration,
BLE shows lower power consumption (lower than 1 mW) over the range of 50 bps to
50 kbps. In addition,
the 433 MHz system is unable to sustain effective
throughputs beyond 50 kbps. As a result, for
applications requiring throughput above 50 kbps,
BLE provides a more practical option than the
evaluated 433 MHz configuration.

When comparing BLE and the 915 MHz system, the
power consumption curves intersect at
approximately 100 kbps, where BLE operating at 20
dBm consumes about 12 mW, comparable to the 915
MHz configuration. Consequently, for low-to-moderate-throughput
ingestible applications---including pH,
temperature, pressure, gas, electrochemical,
vital-sign, and gastric electrophysiology
monitoring---with throughput requirements below
100 kbps, BLE can substantially reduce overall
system power consumption. For example, within the
low-throughput regime from 50 bps to 10 kbps, BLE
achieves approximately one order of magnitude
lower power consumption than the 915 MHz
configuration. In contrast, at target throughputs
above 100 kbps, the 915 MHz Sub-GHz system becomes
more energy efficient than BLE operating at 20
dBm. Moreover, the high-speed mode of the 915 MHz
configuration achieves an effective throughput of
up to 2300 kbps, nearly twice that of BLE
operating under the 2M PHY, highlighting its
advantage in high-throughput scenarios.

When comparing BLE itself operated at 8 dBm and 20
dBm, the 20 dBm configuration incorporates an
external amplifier, resulting in a peak current of
118.73 ± 1.17 mA. In contrast, BLE operating at 8
dBm exhibits a substantially lower peak current of
28.25 ± 0.15 mA. Although the instantaneous power
consumption of the 20 dBm configuration is
theoretically much higher than that of the 8 dBm
configuration, this difference does not translate
into a significant increase in average power
consumption at low throughputs. In the
low-throughput regime (e.g., 50 bps–10 kbps), data
transmission occupies only a small fraction of the
duty cycle, and the FEM remains in a low-power
sleep state for the majority of the time. As a
result, the time-averaged power consumption of BLE
at 20 dBm is comparable to that at 8 dBm, leading
to nearly overlapping power–throughput curves in
this regime. However, at high throughput levels,
the power consumption of BLE operating at 20 dBm
increases substantially.

\subsection{Communication Reliability Evaluation}

\begin{figure}[htbp]
    \centering
    \includegraphics[width=\columnwidth]{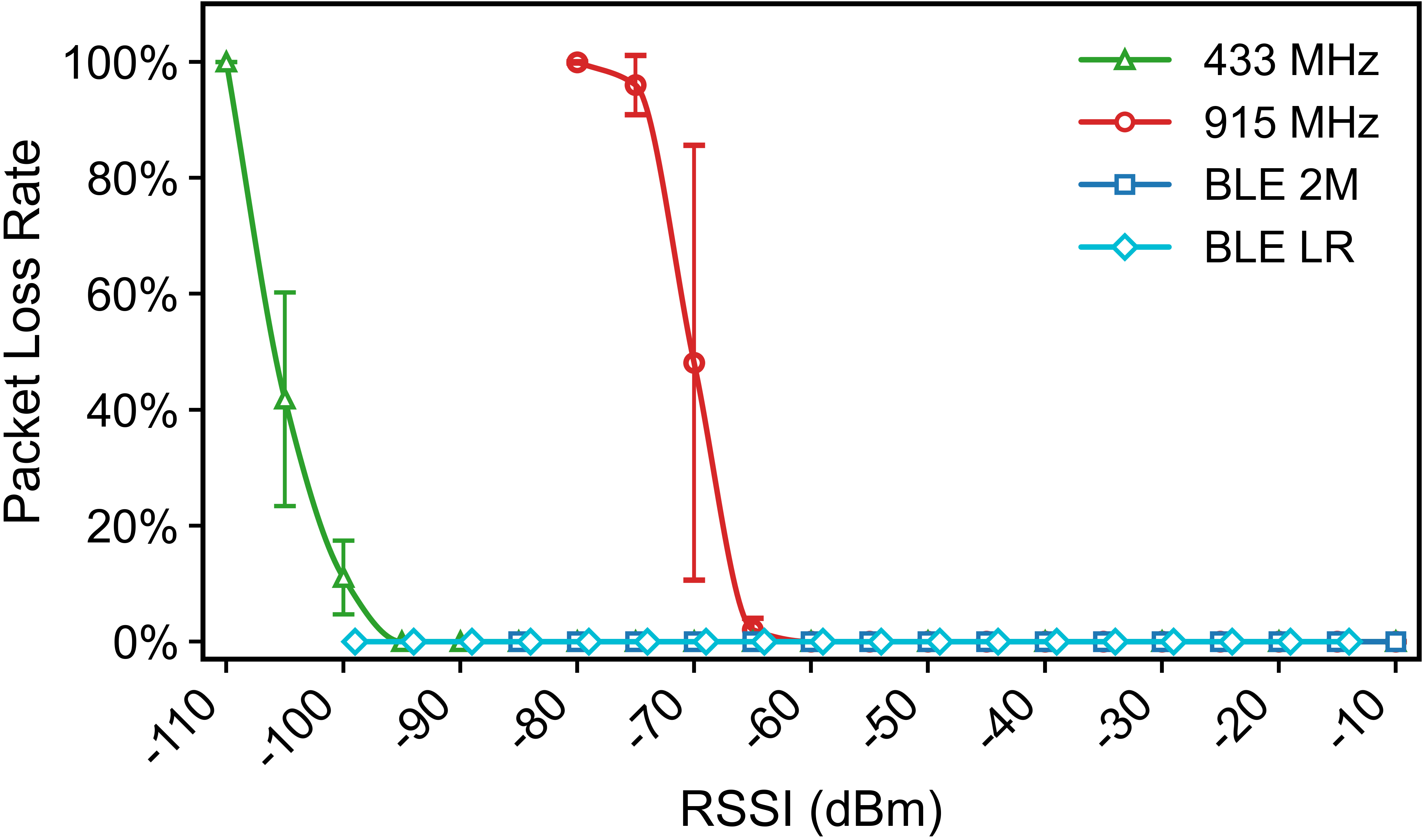}
    \caption{Packet loss rate as a function of RSSI for 433 MHz, 915 MHz,  BLE 2M PHY and  BLE Coded PHY communication.}
    \label{fig:loss_rate}
\end{figure}

Fig. \ref{fig:loss_rate} shows the packet loss
rates of the BLE 2 Mbps PHY, BLE S8 Coded PHY
(LR), and CC1310 operating at 433 MHz and 915 MHz.
For the BLE 2 Mbps PHY, the packet loss rate
remains at 0\% within the connected RSSI range,
namely above -85 dBm. In comparison, the BLE S8 extends reliable packet delivery to
approximately -99 dBm by increasing the link
budget through coding redundancy. However, this
improvement is achieved at the cost of a lower
data rate. The maximum throughput of the BLE S8 was approximately 50 kbps, resulting in higher
energy consumption per bit. Therefore, the BLE 2
Mbps PHY is preferred when the link budget is
sufficient, whereas the BLE S8 is more suitable
for severely attenuated links.

For the CC1310 configurations, packet loss appears
only at low RSSI levels, but the loss rate can
approach 100\% under poor channel conditions. The
433 MHz link begins to lose packets at
approximately -100 dBm, whereas the 915 MHz link
begins to lose packets at around -60 dBm. This
difference is mainly caused by the PHY
configurations used in the two bands. The 433 MHz
configuration uses 2-FSK modulation, in which each
symbol carries 1 bit, and the frequency separation
between symbols is relatively large, providing
higher noise robustness and receiver sensitivity.
In contrast, the 915 MHz high-speed mode uses
8-FSK modulation, in which each symbol carries 3
bits with smaller frequency spacing between
adjacent symbols. This higher-order modulation
improves spectral efficiency but reduces
robustness to noise and frequency offset, leading
to packet loss at a higher RSSI level.

The zero packet loss observed for both BLE PHYs is
attributed to the stop-and-wait automatic repeat
request (ARQ) mechanism at the BLE Link Layer,
which uses the sequence number (SN) and next
expected sequence number (NESN)
\cite{BluetoothCoreSpecV62}. A packet is advanced
only after acknowledgment through an NESN toggle;
otherwise, it is retransmitted, while duplicated
packets with an unchanged SN are discarded. As a
result, BLE provides in-order and duplicate-free
delivery as long as the connection remains active.
In contrast, the evaluated Sub-GHz links do not
provide the same connection-oriented
retransmission and duplicate-suppression behavior
in this configuration, causing packet loss to
increase as RSSI decreases.

Overall, these results indicate that BLE provides
stronger packet delivery reliability than the
evaluated Sub-GHz configurations within its
connection range. This reliability is particularly
important for time-sensitive in vivo sensing or
actuation applications, where continuous data
integrity is required. However, BLE reliability
remains conditional on connection survival; after
supervision timeout and connection termination,
reliable delivery is no longer guaranteed.

\subsection{Bidirectional Communication Evaluation}

\begin{figure}[htbp]
    \centering
    \includegraphics[width=0.8\columnwidth]{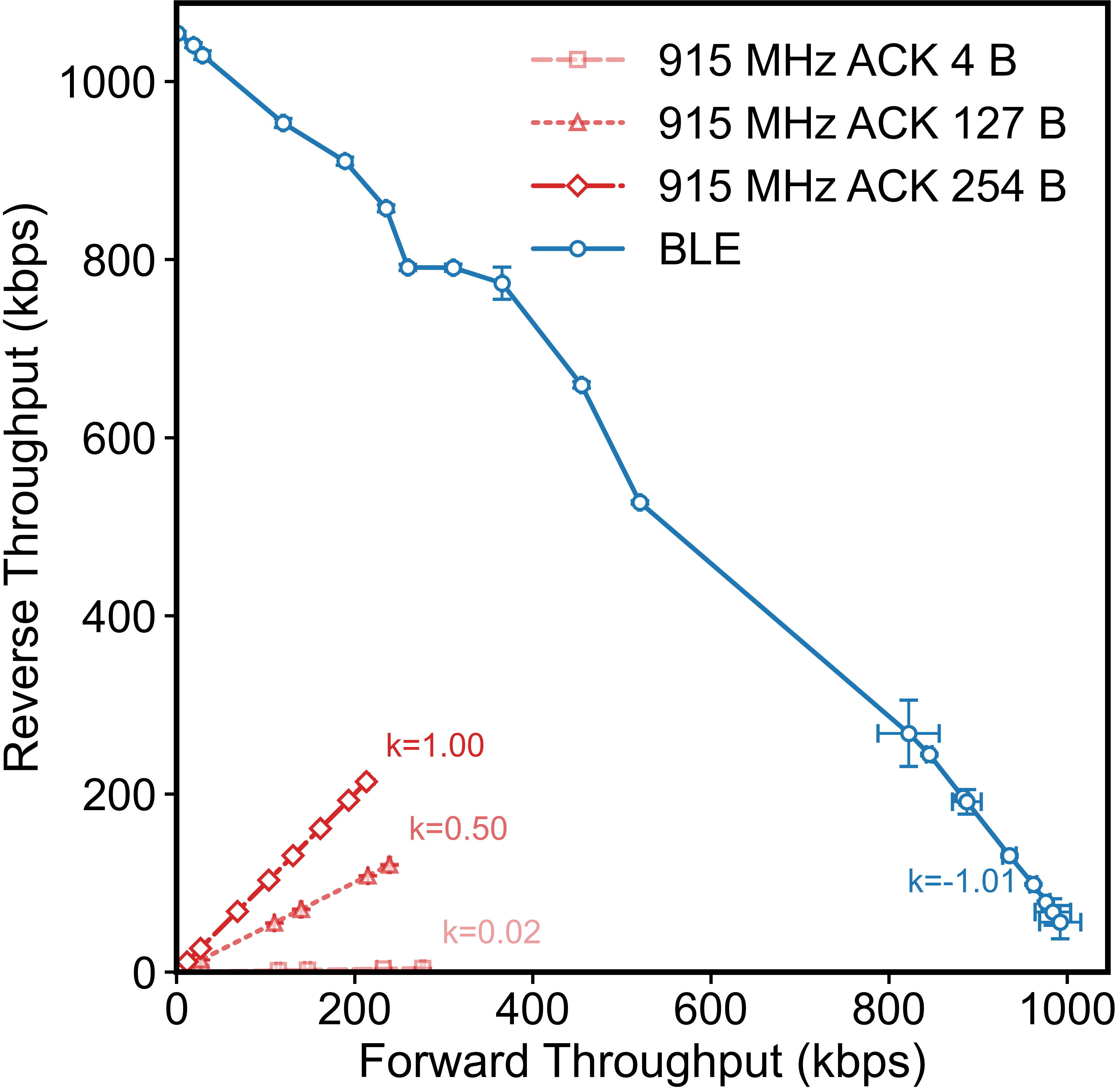}
    \caption{Bidirectional throughput trade-off between forward and reverse transmission for 915 MHz and BLE communication under different ACK payload sizes.}
    \label{fig:bidirectional_tradeoff}
\end{figure}

Ingestible electronics require not only in
vivo-to-ex vivo data transmission but also ex
vivo-to-in vivo command delivery. For example, in
drug delivery systems, external wireless commands
may be used to trigger drug release \cite{Huang2025A}. Similarly,
some advanced gastric-resident devices rely on
wireless commands to initiate device disassembly
\cite{Huang2026Kirigami}. In wireless capsule
endoscopy, control commands may also need to be
received in real time without interrupting the
ongoing video stream
\cite{11252789}. These downlink commands must therefore be delivered effectively and reliably, while
minimally interfering with concurrent uplink
transmission.

Fig. \ref{fig:bidirectional_tradeoff} compares the
bidirectional communication characteristics of BLE
and Sub-GHz links. Owing to its inherent
protocol-level advantages, BLE supports
independent transmission in both forward and
reverse directions. The relationship between the
forward and reverse throughputs follows a
trade-off with a slope of -1, with their combined
throughput remaining approximately 1100 kbps.

The CC1310 4M proprietary PHY used in this study
does not natively provide a bidirectional
communication mechanism. To enable bidirectional
transmission, we implemented the custom
state-machine-based protocol. Although this
protocol enables data transmission in both
directions, it results in a substantial throughput
degradation compared with unidirectional
operation. Under the 8-FSK configuration, the
maximum unidirectional throughput reaches
2300 kbps, whereas the maximum single-direction
throughput decreases to approximately 230 kbps
when bidirectional communication is enabled,
corresponding to a nearly tenfold reduction. This
degradation is primarily caused by repeated TX/RX
state transitions, radio-turnaround latency,
response-frame transmission, and protocol
coordination overhead. Because the lower-rate
2-FSK configuration cannot provide sufficient
throughput under the proposed bidirectional
protocol, it is not included in
Fig.~\ref{fig:bidirectional_tradeoff}.

In addition, the maximum achievable forward and
reverse throughputs are strongly dependent on the
ACK packet length. When the forward and reverse
packets are fully symmetric, the trade-off curve
exhibits a slope of approximately 1. When the
reverse packet length is reduced, the forward
throughput increases, whereas the reverse
throughput decreases accordingly. Moreover,
because the current timing scheme is not aligned
at the protocol layer, reverse transmission must
be triggered by forward packets, preventing
independent control of the two transmission
directions.

The achievable bidirectional throughput admits a
compact closed-form description. Let $B_d$ and
$B_a$ denote the forward DATA and reverse ACK
payload lengths in bytes. The round-trip time
satisfies
\begin{equation}
T_{\mathrm{RTT}} = 2\tau + (B_d + B_a + 30) \times 2~\mu\mathrm{s},
\label{eq:trtt}
\end{equation}
with forward and reverse throughputs
\begin{equation}
\mathrm{TP}_{\mathrm{fwd}} = \frac{8 B_d}{T_{\mathrm{RTT}}}, \qquad
\mathrm{TP}_{\mathrm{rev}} = \frac{8 B_a}{T_{\mathrm{RTT}}}.
\label{eq:tp}
\end{equation}
The second term in Eq.~\eqref{eq:trtt} is the
on-air cost at the 4 Mbps rate, where the extra 30
bytes account for the framing overhead of both
packets (preamble, sync word, length field, and
CRC). The constant $\tau$ captures the per-side
overhead incurred each time the radio switches
direction, which, under the present
implementation, includes the radio-core power-up
and setup, synthesizer re-lock, ACK timeout
configuration, and RTOS task dispatch. Fitting
$\tau$ to the measurements gives $\tau \approx
3.4$ ms, and the resulting model reproduces the
observed throughput within 1.7 percent across ACK
sizes from 4 to 254 bytes.

Overall, Sub-GHz communication requires additional
protocol-level coordination to support reverse
transmission, which inevitably reduces forward
throughput. In contrast, BLE natively supports
independently controlled, forward and reverse
transmission, allowing short reverse commands to
be delivered with minimal impact on forward
throughput.
\subsection{End-to-End Latency Evaluation}

\begin{figure}[htbp]
    \centering
    \includegraphics[width=\linewidth]{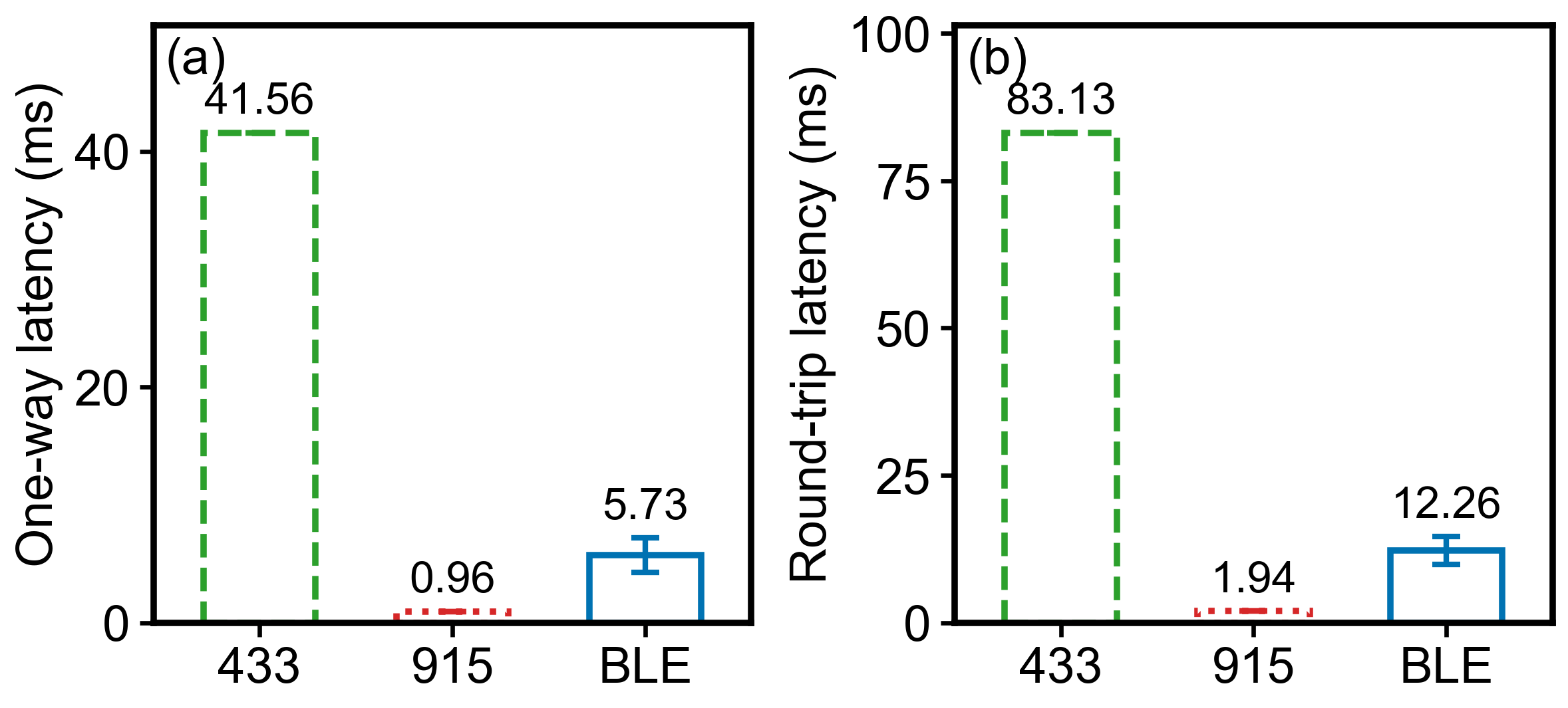}
    \caption{Board-to-board latency comparison for
    the 433 MHz, 915 MHz, and 2.4 GHz BLE
    configurations. (a) One-way latency. (b)
    Round-trip latency.}
    \label{fig:latency_board}
\end{figure}

Fig.~\ref{fig:latency_board} compares the
board-to-board latencies of BLE and the two
Sub-GHz configurations. The 915 MHz high-speed
link achieves the lowest latency, with one-way and
round-trip latencies of 0.96 ms and 1.94 ms,
respectively. BLE requires 5.73 ms one-way and
12.26 ms round-trip, while the 433 MHz
configuration shows the highest latencies of
41.56 ms and 83.13 ms. These results primarily
reflect differences in physical-layer data rate,
particularly the long transmission time of the
244-byte packet over the 50 kbps 433 MHz link.

\begin{figure}[htbp]
    \centering
    \includegraphics[width=\linewidth]{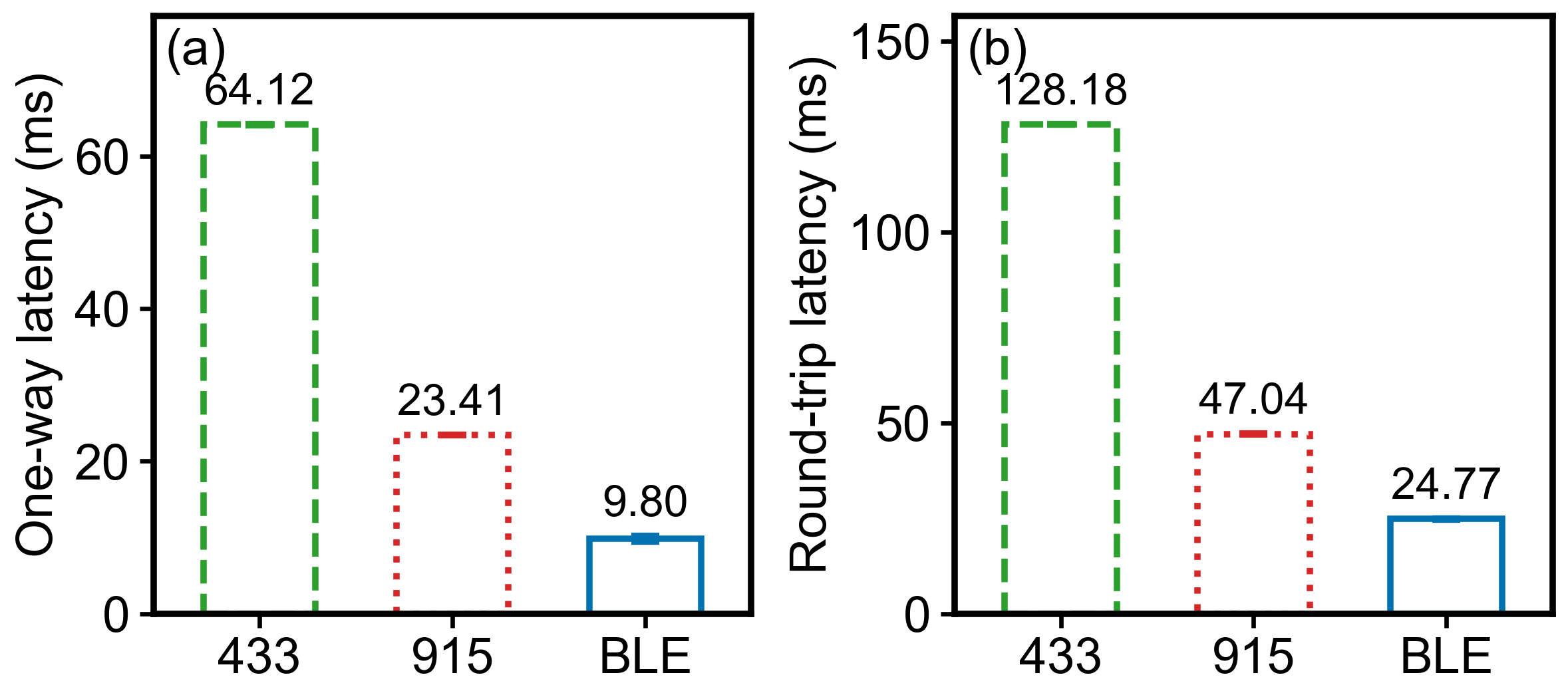}
    \caption{End-to-end latency comparison for the
    433 MHz, 915 MHz, and 2.4 GHz BLE
    configurations. (a) One-way latency. (b)
    Round-trip latency.}
    \label{fig:latency_end_to_end}
\end{figure}

Fig.~\ref{fig:latency_end_to_end} presents the
end-to-end measurements, which additionally
include receiver processing, interface transfer,
and host-side software latency. Although 915 MHz
provides the fastest board-to-board communication,
BLE achieves the lowest end-to-end latency, at
9.80 ms one-way and 24.77 ms round-trip. This is
because BLE communicates directly with the host,
whereas both Sub-GHz configurations require an
external receiver and USB-based forwarding. The
one-way latency increases from 0.96 to 23.41 ms
for 915 MHz and from 41.56 to 64.12 ms for 433
MHz. The similar additional delays of
approximately 22.45 ms and 22.56 ms indicate that
the forwarding and host-processing overhead is
largely independent of the Sub-GHz radio data
rate.

For all configurations, the round-trip latency is
approximately twice the one-way latency,
indicating a relatively symmetric forward and
return path. The 433 MHz configuration
consequently exhibits the highest end-to-end
latency, reaching 64.12 ms one-way and 128.18 ms
round-trip. In contrast, the higher physical-layer
rate of the 915 MHz configuration reduces these
values to 23.41 ms and 47.04 ms. BLE round-trip
latency is approximately 2.14 times its
board-to-board one-way latency and 2.53 times its
end-to-end one-way latency.

Overall, 915 MHz provides the fastest radio-level
communication, while BLE achieves the lowest
practical end-to-end latency because it avoids the
external receiver and USB-forwarding overhead. The
433 MHz link trades substantially higher latency
for its lower-rate Sub-GHz operation.

\subsection{Mechanical Integration and Ecosystem Evaluation}

In ingestible electronics, the antenna occupies a
significant fraction of the available device
volume and strongly constrains mechanical layout, ground
design, and interconnect routing. Consequently,
its form factor plays a critical role in
system-level trade-offs among power storage,
sensing and actuation, packaging, and the
radiating structure itself. Table
\ref{tab:antenna_size} summarizes representative
antenna form factors and dimensions reported for
ingestible devices at 2.4 GHz (BLE), 433 MHz, and
915 MHz. In the surveyed literature, 2.4 GHz (BLE)
ingestible systems typically employ
millimeter-scale chip antennas with a
representative volume from 0.185 to 2.56 mm$^3$
\cite{Stine2023Miniaturized, Chen2025An}. By contrast, Sub-GHz implementations generally require
larger antennas due to the longer wavelength. For
915 MHz systems, the antenna volume is typically
on the order of several tens of cubic millimeters
\cite{You2024An,nadeau_prolonged_2017}, whereas
433 MHz designs predominantly rely on large
helical antennas
\cite{Traverso2023First-in-human, 8094934}, which impose more
stringent constraints on spatial layout within
capsule-based platforms. In this context, a key
practical advantage of BLE lies in its antenna
form factor: the use of compact, modular antennas
minimizes geometric coupling with the capsule
layout, thereby simplifying system integration and
improving manufacturing repeatability.

From a deployment perspective, BLE benefits from a
unified protocol and a well-established ecosystem
that enables seamless interoperability with
consumer smartphones and hospital IT
infrastructures, eliminating the need for custom
receivers or proprietary gateways. This
significantly simplifies deployment, enhances
patient compliance, and supports regulatory
approval and clinical translation. In addition,
BLE offers standardized capabilities for
encryption, device authentication, and
over-the-air firmware updates—features that are
critical for cybersecurity and medical device
lifecycle management but are often less mature or
vendor-specific in Sub-GHz platforms.
Collectively, these ecosystem-level advantages
make BLE a more scalable option for ingestible
electronics.


\begin{table}[htbp]
    \centering
    \caption{Antenna form-factor comparison}
    \label{tab:antenna_size}
    \setlength{\tabcolsep}{10pt}
    \renewcommand{\arraystretch}{1.15}

    \begin{tabular}{lccc}
        \toprule
        Frequency band & Type & Dimensions (mm) & Ref. \\
        \midrule
        \multirow{2}{*}{2.4 GHz (BLE)} 
        & Chip & $3.2 \times 1.6 \times 0.5$ & \cite{Stine2023Miniaturized} \\
        & Chip & $1 \times 0.5 \times 0.37$ & \cite{Chen2025An} \\
        \midrule
        \multirow{2}{*}{433 MHz}       
        & Helical& $\approx\varnothing\,9 \times 7$ & \cite{Traverso2023First-in-human} \\
        & Helical & $\approx\varnothing\,10 \times 15$ & \cite{8094934} \\
        \midrule
        \multirow{2}{*}{915 MHz}       
        & Chip & $6 \times 2 \times 1.08$ & \cite{You2024An} \\
        & Chip & $\approx12 \times 6 \times 1 $& \cite{nadeau_prolonged_2017} \\
        \bottomrule
    \end{tabular}
\end{table}

\section{Conclusion}
This work reevaluated BLE as a wireless
communication strategy for ingestible electronics
by benchmarking it against representative Sub-GHz
links across tissue attenuation, power
consumption, throughput, reliability,
bidirectional communication, latency, antenna
size, and ecosystem compatibility. Although 2.4
GHz BLE experiences stronger tissue-induced
attenuation than 433 MHz and 915 MHz links,
FEM-assisted BLE provides sufficient link-budget
compensation for low- and moderate-throughput
applications. For most ingestible sensing
applications below 100 kbps, BLE consumes less
power, provides protocol-level packet reliability,
supports flexible bidirectional communication, and
achieves lower end-to-end latency than the
evaluated Sub-GHz alternatives. Its compact
antenna form factor and mature ecosystem further
simplify device integration and practical
deployment.

These results show that wireless communication for
ingestible electronics should not be selected
based on tissue attenuation alone, but rather
according to application-specific requirements in
throughput, power, reliability, latency,
bidirectional control, and integration. Properly
configured BLE offers a practical and scalable
solution for most low- and moderate-throughput
ingestible applications, whereas 915 MHz links
remain advantageous for high-throughput scenarios
requiring stronger penetration.

Future work will evaluate BLE communication in
longer-term in vivo studies under diverse
gastrointestinal conditions, including different
capsule orientations, body postures, and tissue
depths. Adaptive TXP control and dynamic PHY
selection will also be integrated to further
improve link robustness and energy efficiency in
practical ingestible sensing and closed-loop
therapeutic systems. In addition, future studies
will examine how to integrate the respective
advantages of BLE and Sub-GHz communication to
enable more efficient in-body wireless links.

\bibliographystyle{IEEEtran}
\bibliography{references}

\end{document}